\documentclass[fleqn,twoside]{article}
\usepackage{amsfonts}

\newcommand{\bls}[1]{\renewcommand{\baselinestretch}{#1}}
\def\noi{\noindent}

\makeatletter

\renewcommand{\thesubsubsection}%
        {\arabic{section}.\arabic{subsection}.\arabic{subsubsection}.}
\renewcommand{\@oddhead}{\raisebox{0pt}[\headheight][0pt]{%
   \vbox{\hbox to\textwidth{\rightmark \hfil \rm \thepage \strut}\hrule}}}
\renewcommand{\@evenhead}{\raisebox{0pt}[\headheight][0pt]{%
   \vbox{\hbox to\textwidth{\thepage \hfil \leftmark \strut}\hrule}}}
\newcommand{\heads}[2]{\markboth{\protect\small\it #1}{\protect\small\it #2}}
\newcommand{\Acknow}[1]{\subsection*{Acknowledgement} #1}
\makeatother

\newcommand{\Title}[1]{\noi {\Large #1} \\}
\newcommand{\Author}[2]{\noi{\large\bf #1}\\[2ex]\noi{\it #2}\\}
\newcommand{\Abstract}[1]{\vskip 2mm \begin{center}
     \parbox{16.4cm}{\small\noi #1} \end{center}\bigskip}
\newcommand{\foom}[1]{\protect\footnotemark[#1]}

\newcommand{\email}[2]{\footnotetext[#1]{e-mail: #2}
	\addtocounter{footnote}{1}}
\def\sect{Sec.\,}

\def\nqq{\hspace{-2em}}
\def\nhq{\hspace{-0.5em}}
\def\nhh{\hspace{-0.3em}}
\def\cm{\hspace{1cm}}
\def\inch{\hspace{1in}}

\def\sequ#1{\setcounter{equation}{#1}}
\def\eq{Eq.\,}
\def\eqs{Eqs.\,}
\def\beq{\begin{equation}}
\def\eeq{\end{equation}}
\def\bear{\begin{eqnarray}}
\def\al{&\nhq}
\def\lal{&&\nqq {}}               % left alignment
\def\bearr{\begin{eqnarray} \lal}
\def\ear{\end{eqnarray}}
\def\earn{\nonumber \end{eqnarray}}

\def\dst{\displaystyle}

\def\nn{\nonumber\\ {}}
\def\nnv{\nonumber\\[5pt] {}}
\def\nnn{\nonumber\\ \lal }
\def\nnnv{\nonumber\\[5pt] \lal }
\def\yy{\\[5pt]}

\def\eql{\al =\al}

\def\eqdef{\stackrel{\rm def}{=}}
\def\e{{\,\rm e}}
\def\d{\partial}

\def\sign{\mathop{\rm sign}\nolimits}
\def\diag{\mathop{\rm diag}\nolimits}
\def\dim{\mathop{\rm dim}\nolimits}
\def\const{{\rm const}}
\def\Half{{\dst\frac{1}{2}}}

\def\then{\quad \Rightarrow \quad}

\newcommand{\vars}[1]{\left\{\begin{array}{ll}#1\end{array}\right.}
\newcommand{\lims}[1]{\mathop{#1}\limits}

\def\DAL{\mathop{\raisebox{3.5pt}{\large\fbox{}}}\nolimits\,}
\newcommand{\Theorem}[2]{\medskip\noi {\bf #1. \ }{\sl #2}\medskip}
\def\deg{\mbox{${}^\circ$}\ }

\def\da{\delta\alpha}
\def\dg{\delta\gamma}

\def\dx{\delta x}

\def\eps{\varepsilon}

\def\o{\omega}

\def\A{{\cal A}}
\def\RR{{\cal R}}
\def\S{{\cal S}}
\def\R{{\mathbb R}}

\def\M{{\mathbb M}\,{}}
\def\Mph{\M_{\rm phys}}
\def\V{{\mathbb V}}
\def\oV{\overline{\V}}
\def\oI{{\overline I}}
\def\od{{\overline d}}
\def\og{{\overline g}}
\def\ox{{\overline x}}
\def\uc{{\underline c}}
\def\vY{\vec Y{}}
\def\vZ{\vec Z{}}
\def\oW{\overline W{}}
\def\oY{\overline Y{}}
\def\oZ{\overline Z{}}
\def\tT{\tilde T{}}
\def\hq{{\hat q}}
\def\hL{{\hat L}\,}

\def\Som{\S_{\o}}
\def\sumn{\sum_{i=1}^{n}}
\def\sumt{\sum_{i=2}^{n}}
\def\sums{\sum_s}
\def\sumo{\sum_{\mu\in\Som}}
\def\umx{u_{\max}}
\def\m{{\rm m}}

\def\Qsq{Q_s^2}
\def\Ysq{Y_s^2}
\def\Yoq{Y_\o^2}
\def\Fei{F_{\e I}}
\def\Fmi{F_{\m I}}

\def\Fs{\raisebox{.2ex}{$\lims{F}_s$}{}}
\def\eqos{\ \stackrel{\rm OS}{=}\ }

\def\ds{ds^2_D}
\def\mn{_{\mu\nu}}
\def\mN{_\mu^\nu}
\def\MN{^{\mu\nu}}
\def\Ref#1{Ref.\,\cite{#1}}
\def\rank{\mathop{\rm rank}\nolimits}

\def\sph{spherically symmetric\ }
\def\bh{black hole}
\def\bhs{black holes}
\def\brane{$p$-brane}
\def\branes{$p$-branes}

\bls{1.03}

\heads{K.A. Bronnikov and V.N. Melnikov}
{p-Brane Black Holes as Stability Islands}

\begin{document}
\thispagestyle{empty}
\vskip 3ex
\begin{flushright}
                                       		  {\bf hep-th/0002200}
\end{flushright}

\bigskip

\Title{P-BRANE BLACK HOLES AS STABILITY ISLANDS}

\Author{K.A. Bronnikov\foom 1 and V.N. Melnikov\foom 2}
{Centre for Gravitation and Fundam. Metrology, VNIIMS,
	3-1 M. Ulyanovoy St., Moscow 117313, Russia;\\
Institute of Gravitation and Cosmology, RUDN,
	6 Miklukho-Maklaya St., Moscow 117198, Russia}

\Abstract
    {In multidimensional gravity with an arbitrary number of internal
     Ricci-flat factor spaces, interacting with electric and  magnetic
     $p$-branes, \sph configurations are considered. It is shown that
     all single-brane \bh\ solutions are stable under \sph perturbations,
     whereas similar solutions possessing naked singularities turn out to
     be catastrophically unstable. The \bh\ stability conclusion is
     extended to some classes of configurations with intersecting branes.
     These results do not depend on the particular composition of
     the $D$-dimensional space-time, on the number of dilatonic scalar
     fields $\varphi^a$ and on the values of their coupling constants
     $\lambda_{sa}$. Some examples from 11-dimensional supergravity are
     considered.  }

%%%%%%%%%%%%%%%%%%%%%%%%%%%%%%%%%%%%%%%%%%%%%%%%%
\email 1 {kb@rgs.mccme.ru; kbron@mail.ru}
\email 2 {melnikov@rgs.phys.msu.su}

\section{Introduction}   % S1

In this paper we continue our studies of multidimensional gravitational
models based on $D$-dimensional Einstein equations with fields of
antisymmetric forms of arbitrary rank (see [1--3]
	%\cite{IM5,bim97,br-jmp}
and references therein) as some low-energy limit of a future unified model
(M-, F- or other type). Our main interest here will be in the stability
properties of multidimensional black-hole (BH) and non-BH solutions with
nonzero fields of forms, associated with charged \branes.
There exist a large number of such solutions in arbitrary
dimensions --- see e.g. [2, 4--9] and references therein.
	%\cite{bim97,bobs,IMmapa,im9901,cots-im,im9910,gm}.
They are important in connection with studies of processes at early stages
of the Universe, counts of micro-states in BH thermodynamics and now
especially due to new developments in M-theory \cite{wit95} related to
the AdS/CFT correspondence \cite{mald}. For recent reviews of this rapidly
developing field see, e.g., \cite{peter,kirits}.

BH stability studies have a long history, of which we will only mention
(more or less arbitrarily) some milestones, concerning \sph backgrounds.
Regge and Wheeler \cite{rw} considered the stability of the Schwarzschild
space-time and developed the formalism of spherical harmonics for metric
perturbations. Vishveshwara \cite{vish} finally proved the linear stability
of Schwarzschild BHs; Moncrief \cite{monc} did the same for
Reissner-Nordstr\"om ones. BHs with a conformally coupled scalar field were
shown to be unstable under \sph perturbations \cite{78}, as well as
minimally coupled scalar field configurations in general relativity
possessing naked singularities \cite{hod}. The monopole degree of freedom
is present there due to the scalar field; it was argued that monopole
perturbations were most likely to be unstable due to the absence of
centrifugal terms in the effective potentials; catastrophic
instabilities were indeed found and it was unnecessary to study other
multipoles. On the other hand, coloured BHs, containing non-Abelian gauge
fields, were shown to be, in general, unstable due to their sphaleronic
degrees of freedom --- see \cite{kanti99} and references therein.
A recent overview of 4-dimensional perturbation studies may be found in
\Ref{koksh}.

For BHs in multidimensional theories of gravity the situation is more
complex since, on the one hand, there emerge new effective scalar fields
(extra-dimension scale factors, sometimes called moduli fields) in the
external space-time, and, on the other, instabilities may be caused by waves
in extra dimensions. Instabilities of the latter kind were indeed found by
Gregory and Laflamme \cite{glaf1,glaf2} for a limited class of neutral and
charged black strings and branes, having a constant internal space scale
factor. Furthermore, it was argued that compactification on a sufficiently
small length scale should prevent the onset of instability, and, moreover,
that extremal black branes are stable \cite{glaf3}. It was concluded that
only very light BHs, whose horizon radii have the same order of magnitude
as their extra dimensions, manifest this form of instability.

It is therefore of interest to inquire whether or not there are other forms
of instability, maybe ``more dangerous'', on more general backgrounds,
containing nontrivial internal space structures and/or several dilatonic
scalars and brane charges. As was previously the case with backgrounds
containing effective scalar fields, it is natural to consider first the
simplest, monopole perturbations.

Earlier we analyzed the stability of static, \sph solutions to the
Einstein-Maxwell-scalar equations with a dilatonic type coupling between
scalar and electromagnetic fields in $D$-dimensional gravity
\cite{br-vuz92,bm-ann}. It was proved there that only BH configurations
were stable under linear \sph perturbations, while non-BH solutions
turned out to be catastrophically unstable. A similar result was obtained
for dilatonic BHs with the inclusion of the Gauss-Bonnet curvature term due
to one-loop quantum corrections \cite{kanti98}.  We will now show that in
the simplest case of a single charged black brane the solution is stable
under linear \sph perturbations, whereas single-brane solutions with naked
singularities are unstable. So the results of \cite{br-vuz92,bm-ann} are
generalized.

We also present a tentative consideration of multi-brane BHs and
conclude that in cases when the perturbation equations decouple,
the stability conclusion is also valid. Two classes of such systems
are indicated, both characterized by certain relations among brane charges,
such that, in terms of \sect 3, the constituent vectors $\vY_s$ form a
single block of a block-orthogonal system (BOS) --- single-block BHs for
short.  Namely, the stability is proved for arbitrary two-brane
single-block BHs and multi-brane single-block BHs with mutually orthogonal
vectors $\vY_s$ (see the details in \sect 6.2).  For many single-block
configurations which do not belong to these classes, the stability can be
proved as well, but their properties require individual studies; see an
example in the Appendix, \eqs (\ref{A.4})--(\ref{A.7}).  There are,
however, numerous multi-brane BHs for which decoupling is impossible and
one may expect that some of them show a new type of instability connected
with mode interaction; a study of these systems is in progress.

The paper is organized as follows. \sect 2 describes the general features
of the field model to be considered. \sect 3 presents some known static
solutions, including BHs, on the basis of the target space $\V$ connected
with dimensional reduction. In \sect 4 a truncated target space $\oV$, more
appropriate for treating the perturbations, is introduced, and
wave equations for perturbations are derived. In \sect 5 the stability
properties of single-brane configurations are deduced, while in \sect 6 the
stability of some multi-brane BHs under \sph perturbations is established.
The Appendix gives some examples from 11-dimensional supergravity.

The word ``stable'' throughout the paper means ``stable under linear \sph
perturbations''.

\section{The model} % S2

     Our starting point is, as in Refs.\,[1--8], the model action for
     $D$-dimensional gravity with several scalar dilatonic fields
     $\varphi^a$ and antisymmetric $n_s$-forms $F_s$:
\beq                                                         \label{2.1}
     S = \int\limits_{\M} d^{D}z \sqrt{|g|} \biggl\{
     \RR [g]
	- \delta_{ab} g^{MN} \d_{M} \varphi^a \d_{N} \varphi^b
                                                    - \sum_{s\in \S}
        \frac{\eta_s}{n_s!} \e^{2 \lambda_{sa} \varphi^a} F_s^2
                  \biggr\},
\eeq
     in a pseudo-Riemannian manifold
     $\M = \R_u \times \M_0 \times \ldots \times \M_n$,
     with factor space dimensions $d_i$, $i=0,\ldots,n$; $\RR$ is the
     scalar curvature. We will assume
     $\M$ to be \sph\nhh, so that the metric is
\bear
     \ds  = g_{MN}dz^M dz^N \eql
               \e^{2\alpha^0} du^2 +                         \label{2.11}
    	             \sum_{i=0}^{n} \e^{2\beta^i} ds_i^2  \nn
   	  \eql
              \e^{2\alpha^0} du^2 + \e^{2\beta^0} d\Omega^2
	    - \e^{2\beta^1} dt^2 +
    	             \sum_{i=2}^{n} \e^{2\beta^i} ds_i^2.
\ear
     Here $u$ is a radial coordinate ranging in $\R_u \subseteq \R$;
     $ds_0^2 = d\Omega^2$ is the metric on a unit $d_0$-dimensional sphere
     $\M_0 = S^{d_0}$; $t \in \M_1 \equiv \R_t$ is time;
     the metrics $g^i = ds_i^2$ of the ``extra" factor spaces
     ($i\geq 2$) are assumed to be Ricci-flat and can have arbitrary
     signatures $\eps_i=\sign g^i$;
     $|g| = |\det g_{MN}|$ and similarly for subspaces;
     $F_s^2 = F_{s,\ M_1 \ldots M_{n_s}} F_s^{M_1 \ldots M_{n_s}}$;
     $\lambda_{sa}$ are coupling constants;
     $\eta_s = \pm 1$ (to be specified later);
     $s \in \S$,  $a\in \A$, where $\S$ and $\A$ are some finite sets.
     The ``scale factors" $\e^{\beta^i}$ and the scalars $\varphi^a$ are
     assumed to depend on $u$ and $t$ only.

     The $F$-forms should be also compatible with spherical symmetry.
     A given $F$-form may have several essentially (non-permutatively)
     different components; such a situation is
     sometimes called ``composite $p$-branes"%
\footnote{There is an exception:  two components,
   having only one noncoinciding index, cannot coexist since in this case
   there emerge nonzero off-block-diagonal components
   of the energy-momentum tensor (EMT) $T_M^N$, while the
   Einstein tensor in the l.h.s. of the Einstein equations is
   block-diagonal.  See more details in \Ref {IM5}.}.
     For convenience, we will nevertheless treat essentially different
     components of the same $F$-form as individual (``elementary") $F$-forms.
     A reformulation to the composite ansatz, if needed, is straightforward.

     Each $n_s$-form
     $F= dA \equiv \d_{[M_1} A_{M_2\ldots M_{n_s}]}dz^{M_1}\ldots
     dz^{M_{n_s}}$ is then associated with a certain subset $I = \{i_1,
     \ldots, i_k \}$ ($i_1 < \ldots < i_k$) of the set of numbers labelling
     the factor spaces:  $\{i\} = I_0 = \{0, \ldots, n \}$.  The forms $F_s$
     are naturally classified as {\it electric\/} ($\Fei$) and {\it
     magnetic\/} ($\Fmi$) ones.  By definition, the potential $A_I$ of an
     electric form $\Fei$ carries the coordinate indices of
     the subspaces $\M_i,\ i\in I$ and is $u$-dependent
     (since only a radial component of the field may be nonzero).
     A magnetic form $\Fmi$ is built as a form dual to a possible electric
     one, and its nonzero components carry coordinate indices of the
     subspaces $\M_i,\ i\in \oI \eqdef I_0 \setminus I$, One can write:
\beq
    n_{\e I} = \rank F_{\e I} = d(I) + 1,\cm
    n_{\m I} = \rank F_{\m I} = D - \rank F_{\e I} = d(\oI)  \label{2.22}
\eeq
    where $d(I) = \sum_{i\in I} d_i$ are the dimensions of the subspaces
    $\M_I = \M_{i_1} \times \ldots \times \M_{i_k}$.
    The index $s$ will be used to jointly describe the two types of forms,
    so that \cite{bim97,bobs}
\beq                                                     \label{2.23}
     \S = \{s\} = \{\e I_s\} \cup \{\m I_s\}.
\eeq

    We will make some more assumptions to assure that all $F$-forms behave
    like genuine electric or magnetic fields in the physical subspace $\Mph
    = \R_u\times \R_t \times \M_0$, namely:
\def\ii{\phantom{i}}
\bear                                                           \label{i}
 {\bf(i)} && \quad     1 \in I_s,\ \ \forall s \quad
     \mbox{(the subspaces $M_{I_s}$ contain the time axis $\R_t$)};\\
								\label{ii}
 {\bf(ii)} &&\quad       0 \not\in I_s,\ \  \forall s \quad
	\mbox{(the branes only ``live'' in extra dimensions)};
			                                 \\    \label{iii}
{\bf(iii)} &&\quad  -T^t_t (F_s) > 0,\ \ \forall s \quad
	\mbox{(the energy density is positive).}
\ear

    By (i), the so-called quasiscalar forms \cite{bim97,bobs} (forms
    with $1 \not\in I_s$, behaving as effective scalar or pseudoscalar
    fields in $\Mph$) are excluded. The reason for adopting (i) is that our
    interest here is mostly in BHs which do not admit nonzero quasiscalar
    forms (the no-hair theorem for brane systems \cite{br-jmp}).

    Assumption (iii) holds if all extra dimensions are spacelike
    ($\eps_i=1$, $i\geq 2$) and in (\ref{2.1}) all $\eta_s=1$.
    In more general models, with arbitrary $\eps_i$, (iii) holds if
\bearr
     \eta_{\e I_s} = - \eps(I_s),\cm
     \eta_{\m I_s}=  \eps(\oI_s), \cm                        \label{2.25}
     		\eps(I) \eqdef \prod_{i\in I}\eps_i.
\ear

     We will consider static configurations and their
     small (linear) time-dependent perturbations. It turns out, however,
     that under the above assumptions the Maxwell-like field
     equations for $F_s$ may be integrated in a general form for their
     arbitrary dependence on $u$ and $t$. Indeed, for an electric $m$-form
     $F_s$ ($s = \e I$, $m= d(I_s)+1$)
     the field equations due to (\ref{2.1})
\beq                                                        \label{Max}
     {\d_u \choose \d_t}\Bigl(
     F_s^{utM_3\ldots M_m}\sqrt{|g|}\e^{2\lambda_{sa}\varphi^a}\Bigr)=0
\eeq
     are easily integrated to give
\beq
     F_s^{utM_3\ldots M_m} = Q_s                             \label{el}
      	\e^{-\alpha^0-\sigma_0-2\lambda_{sa}\varphi^a}
                      \eps^{M_3...M_{d(I)}}/\sqrt{|g_I|}
		\quad \then \quad
      \frac{1}{m!} F_s^2
      	= \eps(I) Q_s^2 \e^{-2\sigma(\oI)-2\lambda_{sa}\varphi^a}.
\eeq
     where $\eps^{...}$ and $\eps_{...}$ are Levi-Civita symbols,
     $|g_I| = \prod_{i\in I}|g^i|$, and $Q_s= \const$ are charges.
     In a similar way, for a magnetic $m$-form $F_s$ ($s= \m I$,
     $m=d(\oI_s)$), the field equations and the Bianchi identities
     $dF_s=0$ lead to
\beq                                                         \label{mag}
     F_{s, M_1... M_{d(\oI)}} = Q_s
		\eps_{M_1...M_{d(\oI)}} \sqrt{|g_\oI|}
	\quad \then \quad
      \frac{1}{m!} F_s^2
      	= \eps(\oI) Q_s^2 \e^{-2\sigma(\oI) + 2\lambda_{sa}\varphi^a}.
\eeq
     We use the notations
\beq
     \sigma_i = \sum_{j=i}^{n} d_j\beta^j (u,t), \cm       \label{sigma}
     \sigma (I) = \sum_{i\in I} d_i\beta^i (u,t).
\eeq

     Evidently, the expressions (\ref{el}) and (\ref{mag}) differ only in the
     signs before $\lambda_{sa}$ and the signature-dependent prefactors
     $\eps$. Due to (\ref{2.25}), their energy-momentum tensors (EMTs)
     coincide up to the replacement $\lambda_{sa}\to -\lambda_{sa}$, and
     their further treatment is quite identical. In what follows we
     therefore mostly speak of electric forms, but the results are
     easily reformulated for any sets of electric and magnetic forms.  We
     also assume that all $Q_s\ne 0$.

\section {Static systems}

\subsection{The target space $\V$} %S3

     Under the above assumptions, the system is well described using the
     so-called $\sigma$ model representation \cite{IM5}), to be briefly
     outlined here as applied to static, \sph systems. This formulation
     can be derived by reducing the action (\ref{2.1}) to the
     $(d_0+1)$-dimensional space $\R_u \times \M_0$.

     As in \cite{br73} and many later papers, we choose the harmonic
     $u$ coordinate ($\nabla^M \nabla_M u = 0$), such that
\beq                                                         \label{3.1}
     \alpha^0 (u)= \sigma_0 (u) \equiv \sum_{i=0}^{n} d_i \beta^i.
\eeq
     Due to (\ref{ii}), the combination ${1\choose 1} + {2\choose 2}$
     of the Einstein equations has a Liouville form and is integrated giving
\bear
     \e^{\beta^0 - \alpha^0} = (d_0-1) s(k, u),\cm             \label{3.8}
      s(k,u) \eqdef \vars{ k^{-1} \sinh ku, \quad & h>0,\\
     			                 u,       & h=0,\\
     			    k^{-1} \sin ku,       & h<0. }
\ear
     where $k$ is an integration constant. With (\ref{3.8}) the
     $D$-dimensional line element may be written in the form
     ($\od \eqdef d_0-1$)
\beq
     \ds= \frac{\e^{-2\sigma_1/\od}}{[\od s(k,u)]^{2/\od}}
     	  \Biggl\{ \frac{du^2}{[\od s(k,u)]^2} + d\Omega^2\Biggr\}
     	      - \e^{2\beta^1} dt^2
	      	    + \sumt \e^{2\beta^i}ds_i^2.              \label{3.10}
\eeq
     The $u$ coordinate is defined for $0 < u < \umx$ where $u=0$
     corresponds to spatial infinity while $\umx$ may be finite or infinite
     depending on the form of a particular solution.

     The remaining set of unknowns ${\beta^i(u),\ \varphi^a (u)}$
     ($i = 1, \ldots, n,\ a\in \A$) can be treated
     as a real-valued vector function $x^A (u)$ (so that
     $\{A\} = \{1,\ldots,n\} \cup \A$) in an $(n+|\A|)$-dimensional vector
     space $\V$ (target space). The field equations for $x^A$
     can be derived from the Toda-like Lagrangian
\bearr                                                      \label{3.12}
     L=G_{AB}x_u^A x_u^B + \sums Q_s^2 \e^{2y_s(u)}
     \equiv
     \sumn d_i(\beta_u^i)^2 + \frac{\sigma_{1,\,u}^2}{d_0-1}
	          + \delta_{ab}\varphi_u^a \varphi_u^b
		  		+ \sums Q_s^2 \e^{2y_s(u)}
\ear
     (the subscript $u$ means $d/du$), with the ``energy" constraint
\beq                                                      \label{3.16}
	E = G_{AB}x_u^A x_u^B - \sums Q_s^2 \e^{2y_s}
	                                =\frac{d_0}{d_0-1}k^2 \sign k.
\eeq
     The nondegenerate symmetric matrix
\beq                                                       \label{3.13}
     (G_{AB})=\pmatrix {
  	       d_id_j/\od + d_i \delta_{ij} &       0      \cr
	          0                         &  \delta_{ab} \cr }
\eeq
     defines a positive-definite metric in $\V$;
     the functions $y_s(u)$ are scalar products:
\beq                                                       \label{3.15}
     y_s = \sigma(I_s) - \lambda_{sa}\varphi^a
	   \equiv Y_{s,A}  x^A,    \cm\
     (Y_{s,A}) = \Bigl(d_i\delta_{iI_s}, \ \  -\lambda_{sa}\Bigr),
\eeq
     where $\delta_{iI} =1$ if $i\in I$ and $\delta_{iI}=0$ otherwise.
     The contravariant components and scalar products of the vectors $\vY_s$
     are found using the matrix $G^{AB}$ inverse to $G_{AB}$:
\bearr                                                      \label{3.18}
     (G^{AB}) = \pmatrix{
	\delta^{ij}/d_i - 1/(D-2) &      0      \cr
	0                         &\delta^{ab}  \cr }, \cm\cm
	(Y_s{}^A) =
  \Bigl(\delta_{iI_s}-\frac{d(I_s)}{D-2}, \quad -\lambda_{sa}\Bigr); \\
\lal  Y_{s,A}Y_{s'}{}^A \equiv \vY_s \vY_{s'}
	                  = d(I_s \cap I_{s'})                \label{3.20}
     			      - \frac{d(I_s)d(I_{s'})}{D-2}
			      + \lambda_{as}\lambda_{as'}.
\ear
     The equations of motion in terms of $\vY_s$ read
\beq
     \ddot{x}{}^A = \sums Q_s^2 Y_s{}^A \e^{2y_s(u)}.         \label{eqm}
\eeq

\subsection{Exact solutions: orthogonal systems (OS)}

     The integrability of the Toda-like system (\ref{3.12}) depends on the
     set of vectors $\vY_s$. In many cases general or special solutions to
     \eqs (\ref{eqm}) are known. Here we will mention the simplest case of
     integrability: a general solution is available if all $\vY_s$ are
     mutually orthogonal in $\V$ \cite{bim97}, that is,
\beq                                                        \label{3.21}
     \vY_s \vY_{s'} = \delta_{ss'} Y_s^2, \cm
	     Y_s^2 =
	d(I)\bigl[1- d(I)/(D-2) \bigr] + \lambda^2_{s} >0
\eeq
     where $\lambda^2_s = \sum_a\lambda^2_{sa}$.
     Then the functions $y_s(u)$ obey the decoupled Liouville equations
     $y_{s,uu} = Q^2_s Y^2_s \e^{2y_s}$, whence
\beq                                       	            \label{3.23}
     \e^{-2y_s(u)} = Q^2_s Y^2_s\, s^2(h_s,\ u+u_s)
\eeq
     where $h_s$ and $u_s$ are integration constants and the function
     $s(.,.)$ has been defined in (\ref{3.8}). For the sought functions
     $x^A(u)$ and the ``conserved energy'' $E$ we then obtain:
\bear                                                        \label{3.24}
     x^A(u) \eql \sums \frac{Y_s{}^A}{\Ysq} y_s(u) + c^A u + \uc^A,
 \\
        E \eql \sums \frac{h_s^2\sign h_s}{\Ysq} + \vec c\,{}^2 \label{3.29}
                     = \frac{d_0}{d_0-1} k^2 \sign k,
\ear
     where the vectors of integration constants $\vec c$ and $\vec\uc$ are
     orthogonal to all $\vY_s$: \ $c^A Y_{s,A} = \uc^A Y_{s,A} = 0$, or
\beq
     c^i d_i\delta_{iI_s} - c^a\lambda_{sa}=0, \inch
     \uc^id_i\delta_{iI_s} - \uc^a\lambda_{sa}=0.     \label{3.25}
\eeq

\subsection{Exact solutions: block-orthogonal systems (BOS)}

     The above OS solutions are general for input parameters
     ($D$, $d_i$, $\vY_s$) satisfying \eq (\ref{3.21}): there is an
     independent charge attached to each (elementary) $F$-form.
     One can, however, obtain special solutions for more general sets of
     input parameters, under less restrictive conditions than (\ref{3.21}).
     Namely, assuming that some of the functions $y_s(u)$ (\ref{3.15})
     coincide, one obtains the so-called BOS solutions \cite{bobs}, where
     the number of independent charges coincides with the number of
     different functions $y_s(u)$.

     Indeed, suppose \cite{bobs} that the set $\S$ splits into several
     non-intersecting non-empty subsets,
\beq                                                      \label{*1}
     \S = \bigcup_{\o}\Som, \cm |\Som|=m(\o),
\eeq
     such that the vectors $\vY_{\mu(\o)}$ ($\mu(\o) \in \Som$)
     form mutually orthogonal subspaces $\V_{\o} \subseteq \V$:
\beq
     \vY_{\mu(\o)} \vY_{\nu(\o')} = 0, \cm \o \ne \o'.    \label{*2}
\eeq
     Then the corresponding result from \cite{bobs} can be formulated as
     follows:

\Theorem{Proposition 1}
    {Let, for each fixed $\o$, all $\vY_\nu \in \V_\o$
     be linearly independent, and let there be a vector
     $\vY_\o = \sumo a_{\mu}\vY_{\mu}$ with $a_\mu > 0$ such that
\beq                                                            \label{*3}
     \vY_\mu \vY_\o = Y_\o^2 \eqdef \vY_\o^2,\cm \forall \mu\in\Som.
\eeq
     Then one has the following solution to the equations of motion
     (\ref{eqm}), (\ref{3.16}):
\bear                                                         \label{*6}
     x^A \eql \sum_\o \frac{Y_\o{}^A}{\Yoq} y_\o(u)+ c^A u + \uc^A,\\
							      \label{*7}
     \e^{-2y_{\o}}
         \eql \hq_{\o} \Yoq s^2 (h_\o, u+u_\o),
     \inch
		\hq_\o \eqdef \sumo Q^2_\mu,  \\
							      \label{*5}
       E \eql \sum_\o \frac{h_\o^2\sign h_\o}{\Yoq} + \vec c\,{}^2
                     			= \frac{d_0}{d_0-1} k^2 \sign k
\ear
     where $h_\o$, $u_\o$, $c^A$ and $\uc^A$ are integration constants;
     $c^A$ and $\uc^A$ are constrained by the orthogonality
     relations (\ref{3.25}) \ {\rm (so that the vectors $\vec c$
     and $\vec{\uc}$ are orthogonal to each individual vector $\vY_s\in\V$).}
     }

     \eqs (\ref{*3}) form a set of linear algebraic equations with respect
     to the ``charge factors'' $a_\nu = Q_\nu^2/\hq_{\o}$, satisfying the
     condition $\sumo a_{\mu}=1$. A solution to (\ref{*3}) for
     given $\vY_{\mu}$ can contain some $a_{\mu}<0$; according to
     \cite{bobs}, this would mean that such a \brane\ is ``quasiscalar'',
     violating the assumption (\ref{i}). Solutions with such branes are
     possible but are rejected here since they do not lead to \bhs.
     Furthermore, if a solution to (\ref{*3}) gives $a_\mu=0$ for some
     $\mu\in \Som$, this means that the block cannot contain such a
     \brane, and then the consideration may be repeated without it.%
\footnote
  {Geometrically, the vector
  $\vY_{\o}$ solving \eqs (\ref{*3}) is the altitude of the pyramid formed by
  the vectors $\vY_{\mu}$, $\mu \in \Som$ with a common origin. The
  condition $a_\mu >0$ means that this altitude is located inside the
  pyramid, while $a_\mu=0$ means that the altitude belongs to one of its
  faces.
  \label{foogeom} }

     The function $y_{\o}(u)$ is equal to $y_{\mu(\o)}(u)=
     Y_{\mu(\o),A}x^A$, which is, due to (\ref{*3}), the same for all
     $\mu\in\Som$. The BOS solution generalizes the OS one, (\ref{3.23}),
     (\ref{3.24}): the latter is restored when each block contains a single
     $F$-form.

     Both kinds of solutions are asymptotically flat, and it is
     natural to normalize the functions $y_s(u)$ and $y_\o(u)$ by the
     condition $y_s(0)=0$ or $y_\o(0)=0$, so that the constants $u_s$ and
     $u_\o$ are directly related to the charges.

     Other solutions to the equations of motion are known, connected with
     Toda chains and Lie algebras \cite{IMmapa, gm, im9901, im9910}.

\subsection {Black-hole solutions}

     Black holes (BHs) are distinguished among other \sph solutions by the
     existence of horizons instead of singularities in the physical
     $d_0+2$-dimension space $\Mph$; the extra dimensions and scalar
     fields are also required to be well-behaved on the horizon to provide
     regularity of the $D$-dimensional metric. Thus BHs are described by
     the above solutions under certain constraints upon the integration
     constants. Namely, for the BOS solution (\ref{*3})--(\ref{*5}),
     requiring that all the scale factors $\e^{\beta^i}$ (except
     $\e^{\beta^1} = \sqrt{|g_{tt}|}$ which should tend to zero) and
     scalars $\varphi^a$ tend to finite limits as $u\to \umx$, we get
     \cite{bim97}:
\bear
     h_{\o} = k >0, \cm \forall\  \o; \inch
     c^A = k \sums Y_{\o}^{-2} Y_{\o}{}^A - k \delta^A_1       \label{5.4}
\ear
     where $A=1$ corresponds to $i=1$ (time). The constraint (\ref{3.29})
     then holds automatically. The value $u = \umx = \infty$ corresponds to
     the horizon. The same condition for the OS solution
     (\ref{3.23})--(\ref{3.25}) is obtained by replacing $\o \mapsto s$.

     Under the asymptotic conditions $\varphi^a \to 0$, $\beta^i \to 0$
     as $u\to 0$, after the transformation
\bearr
     \e^{-2ku} = 1 - \frac{2k}{\od r^\od},\cm \od \eqdef d_0-1  \label{5.5}
\ear
     the metric (\ref{3.10}) for BHs and the corresponding scalar fields
     may be written as
\bearr
    \ds=
     \biggl(\prod_{\o}H_{\o}^{A_{\o}}\biggr)\biggl[-dt^2
    	  \biggl(1-\frac{2k}{\od r^\od}\biggr)\prod_{\o} H_{\o}^{-2/Y_\o^2}
     +
       \biggl(\frac{dr^2}{1-2k/(\od r^\od)} + r^2 d\Omega^2\biggr)
          + \sum_{i=2}^{n} ds_i^2
                       \prod_\o H_{\o}^{A_{\o}^i}\biggr]; \nnnv  \label{5.7}
\cm  A_{\o} \eqdef \frac{2}{\Yoq}
                  \sumo \frac{a_\mu d(I_{\mu})}{D-2}
	                        \eqos \frac 2{\Ysq}\frac{d(I_s)}{D-2}; \cm
     A_{\o}^i \eqdef -\frac{2}{\Yoq}
                  \sumo a_\mu\delta_{iI_\mu}
			   \eqos -\frac{2}{\Ysq} \delta_{iI_s};   \yy  \lal
     \varphi^a = -\sum_{\o} \frac{1}{Y^2_{\o}} \ln H_{\o}
		    		\sumo a_\mu \lambda_{\mu a}
              \eqos -\sums \frac{\lambda_{sa}}{\Ysq} \ln H_s,    \label{phi}
\ear
     where $\eqos$ means ``equal for OS, with $\o\mapsto s$",
     and $H_{\o}$  are harmonic functions in $\R_+ \times S^{d_0}$:
\beq                                                        \label{5.8}
	H_{\o} (r) =  1 + p_{\o}/(\od r^\od), \cm
		   p_{\o} \eqdef \sqrt{k^2 + \hq_\o \Yoq} - k.
\eeq

     The subfamily (\ref{5.4}), (\ref{5.7})--(\ref{5.8}) exhausts all BOS BH
     solutions; the OS ones are obtained in the special case of each block
     $\Som$ consisting of a single element $s$.

     The above relations describe the so-called non-extremal BHs.
     Extremal ones, corresponding to minimum \bh\ mass for given charges
     (the so-called BPS limit), are obtained in the limit $k \to 0$. The
     same solutions follow directly from (\ref{*5})--(\ref{*7}) under
     the conditions $h_{\o} = k = c^A =0$. For $k = 0$, the solution is
     defined in the whole range $r>0$, while $r=0$ in many cases
     corresponds to a naked singularity rather than an event horizon, so
     that we no more deal with a \bh.  However, in many other important
     cases $r=0$ is an event horizon of extremal Reissner-Nordstr\"om type,
     with an AdS near-horizon geometry; some examples are mentioned in the
     Appendix.

     Other families of solutions, mentioned at the end of the previouis
     section, also contain BH subfamilies. The most general BH solutions
     are considered in \Ref {im9910}.

\section{Perturbation equations}

\subsection{Truncated target space $\oV$}

     Consider now nonstatic \sph configurations corresponding to the action
     (\ref{2.1}) with the metric (\ref{2.11}) and all field variables
     depending on $u$ and $t$. As before, we are dealing with
     true electric and magnetic forms $F_s$, so that their $I_s \ni 1$, or
\beq
     I_s = {1} \cup J_s, \cm J_s \subset \{2,\ldots, n\}.  	\label{st1}
\eeq

     As in Refs. \cite{br-vuz92,bm-ann}, it is helpful to
     pass to the Einstein frame in the physical
     $(d_0+2)$-dimensional space-time $\Mph = \R_u\times \M_0 \times \M_1$.
     The action (\ref{2.1}) is then rewritten in terms
     of the metric $g\mn$, the $d_0+2$-dimensional part of $g_{MN}$, and is
     transformed to the Einstein frame in $\Mph$ with the metric
\beq
     \og\mn = \e^{2\sigma_2/d_0} g\mn. 				\label{st2}
\eeq
     The electric $n_s$-forms are re-parametrized as follows:
\bear
     F_{utM_3\ldots M_{n_s}} = \Fs_{ut},   	          	\label{st3}
	\inch
     \frac{1}{n_s!} F_s^2 =
     		\Half \Fs\mn \Fs\MN \e^{-2\sigma(J_s)}
	= - \Qsq \e^{-2\lambda_{sa}\varphi^a},
\ear
     where the indices $M_3, \ldots M_{n_s}$ belong to $J_s$;
     here and henceforth the indices $\mu, \nu$ are raised and lowered using
     the metric $\og\mn$;
     in the last equality the solution (\ref{el}) and the positive
     energy assumption (\ref{iii}) are taken into account.

     The action (\ref{2.1}) is written in terms of $\og\mn$ and $\Fs\mn$ as
     follows (up to a constant prefactor, connected with the volume of
     extra dimensions, and a subtracted total divergence):
\bear
     S \eql \int\limits_{\Mph} d^{d_0+2}z \sqrt{|\og|} \biggl\{
          \RR[\og]
         	-\frac{1}{d_0}(\d \sigma_2)^2 - \sumt d_i(\d\beta^i)^2
							\nnn \inch\inch
	   - \delta_{ab} (\d\varphi^a, \d\varphi^b)
                                          - \Half \sum_{s\in \S}
           \Fs\mn \Fs\MN \e^{2\sigma_2/d_0 -2\sigma(J_s)
			       + 2\lambda_{sa} \varphi^a} \biggr\}\nnv
     \eql \int\limits_{\Mph} d^{d_0+2}z \sqrt{|\og|} \biggl\{
           \RR[\og]
		  - H_{KL}(\d x^K, \d x^L)
         - \Half \sum_{s\in \S}\Fs\mn \Fs\MN \e^{-2Z_{s,K} x^K}\biggr\}
	 							\label{st5}
\ear
     where $(\d f, \d g) = \og\MN \d_{\mu}f \d_{\nu} g$,
     $(\d f)^2 = (\d f, \d f)$; the non-degenerate symmetric matrix
\beq
     (H_{KL})=\pmatrix {                                        \label{st6}
  	       d_i d_j/d_0 + d_i \delta_{ij} &       0      \cr
	          0                          &  \delta_{ab} \cr }
\eeq
     defines a positive-definite metric in the vector space $\oV$
     (truncated target space) parametrized
     by the variables $(x^K) = (\beta^2,\ldots, \beta^n; \varphi^a)$; the
     constant vectors $\oZ_s \in \oV$ are characterized by the components%
\footnote
   {We will use the indices $K,L$ for quantities specified in $\oV$ to
    distinguish them from those in $\V$ where the indices $A,B$ are
    used; vectors in $\oV$ are marked with overbars, those in $\V$ by
    arrows. Scalar products are written as
    $\vY\vZ = G_{AB}Y^A Z^B$ (as before) and  $\oY\oZ = H_{KL}Y^K Z^L$.
    }
\beq                                                           \label{st7}
     (Z_{s,K}) = \Bigl(d_i\delta_{iJ_s} - \frac{d_i}{d_0},
     					\ -\lambda_{sa}\Bigr), \cm
     (Z_s{}^K) = (H^{KL}Z_{s,L})
     	       = \Bigl(\delta_{iJ_s}- \frac{d(I_s)}{D-2},
	       				\ -\lambda_{sa}\Bigr)
\eeq
     where the matrix $(H^{KL})$ is inverse to $(H_{KL})$:
\beq                                                           \label{st8}
     (H^{KL}) = \pmatrix{
			\delta^{ij}/d_i - 1/(D-2) &      0      \cr
			0                         &\delta^{ab}  \cr }.
\eeq
     The truncated target space $\oV$ may be considered as
     the hyperplane $x^1 = - \sigma_2/d_0$ in $\V$, with the metric $H_{KL}$
     induced by $G_{AB}$ (\ref{3.13}). The components $H^{KL}$ turn out to
     be the same as $G^{AB}$ for $i\ne 1$; the components $Y_s{}^A$ and
     $Z_s{}^K$ coincide in the same manner. It is easy to find that for
     vectors whose $I_s$ satisfy (\ref{st1}) (which is always the case
     in the present paper),
\beq
     \vY_s \vY_{s'} = \oZ_s \oZ_{s'} + \frac{d_0-1}{d_0},      \label{st9}
\eeq
     whence it follows that, first, when different $\vY_s$ are mutually
     orthogonal in $\V$, the corresponding $\oZ_s$ are never mutually
     orthogonal in $\oV$; second, for any $\vY_s$ whose $I_s\ni 1$
     one has
\beq
     \vY_s \vY_s \equiv \Ysq > \frac{d_0-1}{d_0}.              \label{st10}
\eeq

\subsection {Wave equations}

     The action (\ref{st5}) may be used to obtain the equations governing
     small \sph perturbations of static solutions. The metric
     (\ref{st2}) in $\Mph$ will be written in the form
\beq                                                         \label{st11}
     ds^2_{\rm E} = \og\mn dz^\mu dz^\nu
         = \e^{2\alpha} du^2 + \e^{2\beta} d\Omega^2 - \e^{2\gamma} dt^2
\eeq
     where ``E" stands for the Einstein frame and
     $\alpha(u,t),\ \beta(u,t),\ \gamma(u,t)$ are connected with the
     corresponding quantities from (\ref{2.11}) as follows:
\beq
     \alpha = \alpha^0 + \sigma_2/d_0, \cm
     \beta  = \beta^0 + \sigma_2/d_0,  \cm
     \gamma = \beta^1 + \sigma_2/d_0.                        \label{st12}
\eeq
     Since the field equations for the $F$-forms have been integrated ---
     see (\ref{el}) and (\ref{st3}), the remaining unknowns are
     $\alpha,\beta,\gamma$ and $x^K$ (that is, $\beta^i,\ i\geq 2$, and
     $\varphi^a$). In what follows we will write
\[
     \alpha (u,t) = \alpha (u) + \da (u,t),
\]
     where $\da$ is a small perturbation, and similarly for other
     unknowns. We accordingly preserve only terms linear in $\da$ and
     similar quantities and in time derivatives. The field equations may be
     written in the form
\bearr                                                         \label{S}
     \DAL [\og] \,x^K = \sums Z_s{}^K \Qsq \e^{-2d_0\beta + 2\oZ_s \ox},
		\\ \lal
     \RR\mN = T\mN - \frac{1}{d_0}\delta\mN T^\lambda_\lambda \eqdef \tT\mN
                                                               \label{E}
\ear
     where $\DAL[\og] = \og\MN \nabla_\mu \nabla_\nu$ is the D'Alembert
     operator, while for the nonzero components of the
     EMT $T\mN$ corresponding to (\ref{st5}) one has (no summing in $\mu$)
\bearr
     \tT^\mu_\mu =                                          \label{EMT}
		\e^{-2\alpha} H_{KL} x_u^K x_u^L
         				\diag(0,\ 1,\ 0, \ldots, 0)
     - \sums \Qsq \e^{-2\alpha + 2\gamma + 2\oZ_s \ox}
		\diag\biggl(
1-\frac{1}{d_0},\ -\frac{1}{d_0},\ -\frac{1}{d_0}\ \ldots,\ -\frac{1}{d_0}
				     \biggr),  \nnn
     \tT_{ut} = H_{KL} x_u^K x_t^L
\ear
     where $x_u= \d_u x$ and $x_t = \d_t x$; the first and second
     places under the symbol ``diag'' belong to $t$ and $u$, respectively.

     As in our previous papers on stability, we use the coordinate
     freedom in the perturbed space-time and put
\beq
	\delta\beta \equiv 0                              \label{gauge}
\eeq
     but preserve the harmonic $u$ coordinate condition in the unperturbed
     (static) space-time%
\footnote{This coordinate is harmonic for both metrics (\ref{2.11}) and
   (\ref{st11}); in the latter the coordinate condition has the form
   $\alpha = d_0 \beta + \gamma$.}.
     Then \eqs (\ref{S}) and (\ref{E}) give
\bear                                                      \label{S1}
     \hL \dx^K + x^K_u(\dg_u-\da_u)-2x^K_{uu}\da \eql
         2\sums \Qsq \e^{2\gamma + 2\oZ_s \ox}Z_s{}^K Z_{s,L}\dx^L;  \cm
     \hL \eqdef  -\e^{2d_0\beta}\d_{tt} + \d_{uu};
      	 \\                                                \label{E1}
     	 d_0 \da_t \eql \ox_u \delta\ox_t;  \yy            \label{E2}
   	 d_0 \beta_u (\da_u - \dg_u)
				\eql  2\ox_{uu}\delta\ox -2\beta_{uu}\da,
\ear
     where (\ref{S1}) follows from the
     $\bigl({}_{ut}\bigr)$ component of (\ref{E}) and (\ref{E2}) from
     one of the angular components of (\ref{E}); we have also used the
     equations valid for static systems, in particular, (\ref{eqm}), where,
     according to the definitions of $\vY$ and $\oZ$,
     $y_s(u) = \vY_s\,\vec x =  \gamma + \oZ_s \ox$.
     Integrating (\ref{E1}) in $t$ and omitting the emerging arbitrary
     function of $u$ (since we neglect static perturbations), we obtain
\beq
      d_0 \da = \ox_u \delta\ox.                           \label{A}
\eeq
     Substituting $\da$ from (\ref{A}) and $\da_u - \dg_u$ from (\ref{E2})
     into (\ref{S1}), we finally arrive at the set of wave equations for the
     dynamical degrees of freedom in our system, represented by $\dx^K$:
\bearr
     \hat L \dx^K = 2P^K{}_L \dx^L,
     		\cm
     P^K{}_L = \frac{1}{d_0}\frac{d}{du}
     	\biggl(\frac{x_u^K x_{L, u}}{\beta_u}\biggr)
		+ \sums
		     \Qsq \e^{2y_s(u)} Z_s^K Z_{s,L}.         \label{W}
\ear

     The stability problem is now reduced to a boundary-value problem for
     $\dx^K (u,t)$. Namely, if there exists a nontrivial solution to \eqs
     (\ref{W}) satisfying some physically reasonable conditions at the ends
     of the range of $u$, such that $|\dx^K|$ (at least some of them) grow
     unboundedly with $t$, then the static system is unstable. Otherwise it
     is stable in the linear approximation.

     The condition at infinity, $u=0$, is evident: the perturbations must
     vanish,
\beq
	  dx^K \to 0 \cm {\rm as}\quad u\to 0.                \label{BC1}
\eeq
     It is less evident at $u=\umx$ since some of the
     background static solutions are singular there.
     As in Refs.\,\cite{hod,bm-ann} and others, dealing
     with minimally coupled or dilatonic scalar fields, we will use the
     minimal requirement providing the validity of the perturbation scheme,
     namely
\beq
     	   |\dx^K/x^K|  < \infty.                             \label{BC2}
\eeq
     When the background is regular, this condition requires that
     the perturbation should be regular as well.

\section{Stability properties of single-brane solutions}

\subsection {Decoupling cases}

     \eqs (\ref{W}) in general do not decouple. Even in the simplest case
     when there is only one antisymmetric form $F$ (that is, one
     $p$-brane), so that \eqs (\ref{3.23})--(\ref{3.25}) yield the general
     static solution to the field equations, \eqs(\ref{W}) contain various
     linear combinations of $\dx^K$ with $u$-dependent coefficients.

     There is, however, an important case when \eqs (\ref{W}) do
     decouple for any configuration of $\M$ with the metric (\ref{2.11}),
     namely, the single-brane solution (\ref{3.23})--(\ref{3.25}) under the
     condition that the vector $\overline{c} = (c^K)$ is parallel to
     $\oZ$ in $\oV$:
\footnote
  {Curiously, they are parallel in $\oV$ although the corresponding vectors
  $\vec c$ and $\vY$ are mutually orthogonal in the surrounding target space
  $\V$. For a clear picture, imagine two vectors in 3-dimensional space
  whose projections onto a plane lie on the same ray.}
\beq
	  c^K = BY^K /Y^2, \cm B = \const                      \label{B}
\eeq
     (here and henceforth in this section we omit the index $s$ since, by
     our assumption, it takes only one value).
     This condition is automatically valid for the case of utmost interest,
     BHs with one $p$-brane (``a black $p$-brane''), which, by (\ref{5.4}),
     corresponds to $B=k=h \geq 0$.

     Due to the collinearity condition (\ref{B}) and the constraint
     (\ref{3.29}), the constants are now connected by the relation
\beq
     N'(h^2 \sign h - B^2) = k^2 \sign k - B^2              \label{B2}
\eeq
     with $N'= Y^{-2} (d_0-1)/d_0 < 1$. It turns out that,
     besides BHs, the condition (\ref{B}) is satisfied for some singular
     solutions whose behaviour is quite generic for the system under study:
\begin{enumerate}
\item
     $k>0$, $h\geq 0$, such that $\umx = \infty$ and
     a singularity at the centre of symmetry is attractive at least in
     terms of the metric (\ref{st11}), $\e^{\gamma} \to 0$;
\item
     $h<0$, so that the solution behaviour is determined by the function
     $s(h,u+u_1) = h^{-1}\sin h(u+u_1)$ in (\ref{3.23}) where
     $u_1=\const \in (0, \pi/|h|)$. In this case the central singularity is
     repulsive, $\e^{\gamma} \to \infty$, of Reissner-Nordstr\"om type.
\end{enumerate}

     Due to (\ref{B}), \eqs (\ref{W}) take the form
\bear                                                          \label{W1}
     \hat L \dx^K = 2Z^K
   \biggl[\frac{1}{d_0}\Bigl(\frac{f_u^2}{\beta_u}\Bigr)_u - f_{uu}\biggr]
			(\oZ \delta\ox),
	\cm\cm
     f(u) \eqdef -\frac{y + Bu}{Y^2}
\ear
     with $y(u)$ determined by (\ref{3.23}); the area function $\beta$
     has the form
\beq
     \beta = -\frac{1}{d_0-1}
     \biggl\{ \ln [(d_0-1)s(k,u)] +\frac{1}{N'}f(u)-Bu +\const\biggr\}
							       \label{beta}
\eeq
     where the value of the constant is inessential.

     Since $\oV$ is an $l$-dimensional Euclidean space ($l = n-1+|\A|$),
     there are $l-1$ linearly independent vectors $\oZ_\bot$ such that
     $\oZ_\bot\oZ = 0$. Therefore the set of wave equations (\ref{W1})
     decouples into one equation for $\oZ\delta \ox$ and $l-1$ equations for
     different $\oZ_\bot\delta\ox$:
\bear                                                          \label{W2}
     \hL (\oZ \delta\ox) \eql U(u) (\oZ \delta\ox), \inch
		      U(u) = 2Z^2
     \biggl[\frac{1}{d_0}\Bigl(\frac{f_u^2}{\beta_u}\Bigr)_u -f_{uu}\biggr];
			     \\
     \hL (\oZ_\bot \delta\ox) \eql 0.                     \label{W3}
\ear

     The static nature of the background solution makes it possible to
     separate the variables:
\beq
     \oZ \delta \ox  = \psi(u) \e^{\Omega t}, \inch
     \oZ_\bot \delta \ox  = \psi'(u) \e^{\Omega' t},           \label{psi}
\eeq
     so that \eqs (\ref{W2}) and (\ref{W3}) lead to
\bear
     \psi_{uu} \eql [\e^{2d_0\beta}\Omega^2 + U(u)] \psi,      \label{Om}\\
     \psi'_{uu} \eql \e^{2d_0\beta}{\Omega'}^2 \psi'.          \label{Om'}
\ear
     The existence of an admissible solution of any of these equations with
     a real value of $\Omega$ or $\Omega'$ would mean that the perturbation
     can grow exponentially with time, hence the instability.

     It is hard to solve \eqs (\ref{Om}), (\ref{Om'}) in their full
     range but it is rather easy to assess the asymptotic behaviour of
     their solutions near $u=0$ and $\umx$, and this will be sufficient for
     making stability conclusions.

     In particular, for $u\to 0$, which
     corresponds to spatial infinity, one has
\beq                                                          \label{as}
     U(u) \to 0\cm {\rm and} \cm
     \e^{d_0\beta} \approx c_0u^{-d_0/\od}, \cm c_0= \od\,{}^{d_0/\od}.
\eeq
     The general asymptotic form of solutions to (\ref{Om}) and (\ref{Om'})
     at small $u$ for all cases under study may be written as follows:
\beq
     \psi\ ({\rm or}\ \psi')                                  \label{psias}
     = u^{d_0/(2\od)} \biggl[
		c_1 \exp(c_0 \od \,\Omega u^{-1/\od})
	      +	c_2 \exp(-c_0 \od\,\Omega u^{-1/\od}) \biggr],
     \cm c_1,\ c_2 = \const.
\eeq
     The boundary condition (\ref{BC1}) then requires that in
     (\ref{psias}) $c_1=0$, and it remains to look at the other end of the
     $u$ range, $u\to \umx$.

\subsection{Instability of naked singularities}

     Consider Case 1 of the previous subsection, a ``scalar type''
     singularity. As $u\to\infty$, the relevant functions
     of the static solution behave as follows:
\beq
	y = hu +O(1),  \cm \beta \approx                       \label{ctr}
		-\frac{u}{Y^2 d_0} \frac{N_1-1}{h+k} (k-B)(h-B)
\eeq
     where $N_1 = d_0 Y^2/\od >1$. Since, due to (\ref{B2}), in the present
     case
\[
     |B| > k \ \Longleftrightarrow \ |B| >h  \cm {\rm and} \cm
     |B| < k \ \Longleftrightarrow \ |B| <h,
\]
     one sees that $\e^{\beta(u)} \to 0$ exponentially. The same happens to
     $U(u)$, therefore the asymptotic form of (\ref{Om}) or (\ref{Om'}) is
     simply $\psi_{uu} =0$ or $\psi'_{uu} =0$, so that
\beq
     \psi\ ({\rm or}\ \psi') = c_3 u + c_4                  \label{psi-c}
\eeq
     with constants $c_3$ and $c_4$.
     On the other hand, the background functions $x^K$ also behave
     as $\const \cdot u$ as $u\to\infty$, therefore the second boundary
     condition (\ref{BC2}) is satisfied for any solution (\ref{psi-c}),
     including the one joining the solution (\ref{psias}) with $c_1=0$ at
     small $u$. We conclude that there are growing modes of perturbations
     for any $\Omega$, hence the singular solution is catastrophically
     unstable.

     In Case 2, $h<0$, we have $\umx = \pi/|h| - u_1 <\infty$ and the
     relevant functions in the static solution approach $\umx$ in the
     following way:
\beq
     y(u) \approx - \ln (|h|\Delta u), \cm                    \label{ctr2}
     x^K \approx -\frac{Z^K}{Y^2} \ln(|h|\Delta u), \cm
     \beta \approx \frac{1}{d_0 Y^2} \ln \Delta u \to -\infty
\eeq
     where $\Delta u = \umx - u$.
     One can make sure that $U(u)$ does not
     affect the asymptotic behaviour of solutions to \eq (\ref{Om}) as
     $u\to\umx$ as compared with that of \eq (\ref{Om'}), and for both one
     can write:
\beq
     \psi\ ({\rm or}\ \psi') = c_5 + c_6 \Delta u          \label{psi-c2}
\eeq
     while the condition (\ref{BC2}) only requires
     $|\dx^K/\ln \Delta u| < \infty$.
     Thus the solution satisfies (\ref{BC2}) for any choice of the
     constants $c_5, c_6$, and, as in Case 1, this leads to the instability
     of the background singular solution.

\subsection{Stability of \bhs}

     In the BH case it is again hard to solve \eqs (\ref{Om}),
     (\ref{Om'}), but for our purpose it is sufficient to note that, to
     realize an instability, a solution should begin with a zero value at
     $u=0$ and tend to a finite limit as $u\to \infty$. This is evidently
     impossible for a solution to (\ref{Om'}) since $\psi'_{uu}/\psi' >0$. We
     conclude that at least the $\psi'$ modes of BH perturbations are
     stable. The same reasoning works for the $\psi$ mode provided
     $U(u) \geq 0$ for all $u>0$.  Let us pass to the
     variable $R = r^{\od}/\od$ in the expression for $U$ in (\ref{W2}),
     so that $0<u<\infty$ corresponds to $\infty > R > 2k$:
\bear
     U  \eql  \frac{2Z^2 pR(R-2k)}{Y^2\,(R+p)^2}
     	\biggl\{                                               \label{U}
		 2k + p\biggl[1- \frac{N'R^2}{(R+p')^2}\biggr]
	  + pN'\frac{4kR +2k(p+p') + pp'}{(R+p')^2}
     						\biggr\}, \nnn \cm
     N' \eqdef \frac{1}{Y^2}\frac{d_0-1}{d_0} < 1, \cm
     p' \eqdef p (1-N'),
\ear
     where we have used the explicit form of the single-brane BH solution
     (\ref{3.23})--(\ref{5.4}) and the substitution (\ref{5.5}) with
     $R=r^\od/\od$, replacing $Y_s\to Y$, $p_s\to p$, $u_s\to u_1$.
     Note that $N' <1$ due to (\ref{st10}), so that, in particular, $p'>0$.
     The expression (\ref{U}) is manifestly positive for $\infty > R > 2k$,
     therefore the $\psi$ mode also does not lead to an instability.
     Thus linear stability of all single-brane BH solutions under
     \sph perturbations has been established.

     Our consideration did not apply to extremal BHs since in this case
     the behaviour of the background functions $x^K(u)$ is
     generically singular as $u\to\infty$ ($R = 1/u \to 0$):
\beq
     x^K = - \sums \frac{Z_s^K}{Y^2} \ln\biggl(1+\frac{u}{u_s}\biggr)
								\label{EBH}
\eeq
     and so there is no reason to require $|\dx^K|<\infty$. In some
     cases it is regular (see \sect{3.2} and examples in the Appendix).
     One can see, however, that again, as $u\to \infty$, \eqs (\ref{Om})
     and (\ref{Om'}) for a single-brane extremal BH take the form $\psi_{uu}
     =0$; the linearly growing solution is discarded since it grows faster
     than $x^K$ in (\ref{EBH}), so we are left with a constant and have to
     require $|\psi < \infty|$ for both regular and singular backgrounds.
     Then the same reasoning with $\psi_{uu}/\psi >0$ makes us conclude that
     such allowed solutions with $\Omega>0$ do not exist and extremal BHs are
     stable as well.  Indeed, an explicit form of $U(u)$ is
\beq                                                             \label{Uu}
     U(u) = \frac{2Z^2}{Y^2(u+u_1)^2}
	    \biggl[ 1 + N'\frac{-u_1^2 + N''u^2}{(u_1 + N''u)^2}\biggr]
\eeq
     where $N'<1$ was defined in (\ref{U}) and $N''= 1-N'$. The reasoning
     works since $U > 0$ for all $u>0$ and $U\to 0$ as $u\to\infty$.

     We can now formulate the following result, to be used in the further
     consideration:

\Theorem{Proposition 2}{If a decoupled linear perturbation mode $\xi$ of a
     static, \sph BH solution obeys the equation
\beq
	\hL \xi = U(u)\xi                                        \label{WW}
\eeq
     with $U(u)\geq 0$ (including the case $U\equiv 0$), this mode is
     stable.
     }

\section {Some \bhs\ with multiple branes}

\subsection{Two-brane \bhs}

     We have seen that one-brane singular background solutions are
     catastrophically unstable; we would not like to treat more complex
     singular solutions since there is no reason to believe that
     interaction of modes can prevent the instability. We instead consider
     some multi-brane BH solutions for which the perturbation equations
     decouple and show that they are stable.

     Suppose there is a BH background solution (\ref{5.4})--(\ref{5.8})
     with two branes, so that $s$ takes two values, $s=1,2$. The
     solution is characterized by two charges $Q_s$, two vectors
     $\vY_s\in\V$ and their counterparts $\oZ_s\in\oV$, which we assume
     to be non-collinear (if they are collinear, the consideration
     simplifies and the result is the same as for a single brane).

     The matrix $P^K{}_L$ in the perturbation equations (\ref{W}) may be
     written in the form
\beq
     P^K{}_L = \sums Q_s^2 \e^{2y_s(u)} Z_s^K Z_{s,L}
	    + \sum_{ss'} Z_s^K Z_{s',L} f_{ss'}(u),            \label{2-1}
\eeq
     with
\beq                                                           \label{2-2}
     f_{ss'}(u) = \frac{1}{Y_s^2 Y_{s'}^2}
	   \biggl[\frac{(y_{s,u}+k)(y_{s',u}+k)}{d_0\beta_u}\biggr]_u.
\eeq

     Just as in the one-brane case, one easily separates the ``transversal''
     degrees of freedom: for vectors $\oZ_\bot \in \oV$ such that
     $\oZ_\bot\oZ_s =0$ (they fill a $(\dim \oV -2)$-dimensional plane), the
	function $\xi = \oZ_\bot\delta\ox$ obeys the wave equation (\ref{WW})
	with $U\equiv 0$.

     However, \eqs (\ref{W}) with the matrix (\ref{2-1}) in general do not
     decouple.  An exception is the special case when the two
     functions $y_s$ coincide,
\beq
     y_1 = y_2 = y(u) = k \sinh (ku_1)\big/ \sinh [k(u+u_1)]    \label{2-3}
\eeq
     although the vectors $\vY_s$ are different; in the BOS terminology
     (\sect 3.3) the two vectors $\vY_s$ form a block and in our case this
     single block exhausts the whole system.  If we suppose for simplicity
     that the norms of $\vY_s$ coincide, $Y_1^2 = Y_2^2 = Y^2$, then the
     charges coincide as well, $Q_1^2 = Q_2^2 = Q^2$, and one obtains for
     the two modes $\xi_{\pm} = (oZ_1 \pm \oZ_2)\dx$:
\bear                                                       \label{2-4}
     \hL \xi_+ \eql
     	   2Z^2 (1+\cos \theta)\left[Q^2 \e^{2y} + 2F\right]\xi_+;\\
     \hL \xi_- \eql
     	   2Z^2 Q^2 \e^{2y} (1-\cos \theta) \xi_-           \label{2-5}
\ear
     where $Z^2 = Y^2 - (d_0-1)/d_0$ (see (\ref{st9}), $\theta$ is the angle
     between the vectors $\oZ_1$ and $\oZ_2$ in $\oV$ and $F$ is the
     function (\ref{2-2}) with $Y^2_1=Y^2_2$ and $y_1=y_2=y$.

     A direct substitution of $y$ and $\beta_u$ into (\ref{2-4}) shows, as
     before, that the coefficient by $\dx_+$ at the r.h.s. is nonnegative;
     for $\dx_-$ in (\ref{2-5}) this is manifestly so. Hence the previous
     reasoning works and we conclude that such BHs (including extremal ones)
     with two branes are stable.

     The case $Y_1^2 \ne Y_2^2$ is covered in the next section.

\subsection{Single-block \bhs}

     A natural question arises, whether or not the stability conclusion
     of the previous section extends to an arbitrary multi-brane BH
     described by a single function $y(u)$, in other words, to any
     single-block BH. Note that any set of linearly independent vectors
     $\vY_s$ may be treated as a BOS-block, hence a special static solution
     of this kind (and hence a BH solution) may always be obtained;
     the only restriction is $a_\mu>0$ for the charge factors
     obeying the consistency conditions (\ref{*3}).

     Consider such a system: let there be a BOS BH solution with $m$
     linearly independent vectors $\vY_s\in \V$, $s\in \S=\Som$, and the
     charge factors $a_s$ satisfy (\ref{*3}). The following relations are
     valid:
\beq
     \vY_{\o} = \sums a_s \vY_s; \cm           		 \label{m1}
     \vY_s \vY_{\o} = Y_{\o}^2, \quad \forall s;    \cm   \sums a_s =1.
\eeq
     It is easy to see that, due to (\ref{st9}), similar relations hold for
     the corresponding vectors $\oZ_s\in \oV$:
\beq
     \oZ_{\o} = \sums a_s \oZ_s \quad \forall s;
     			\cm \oZ_s \oZ_{\o} = \oZ_{\o}^2. 	 \label{m2}
\eeq
     For certainty we suppose that $\oZ_s$ are linearly independent;
     if they are not, the consideration is slightly modified without
     changing the results.

     The wave equations (\ref{W}) take the form
\beq                                                             \label{Wm}
     \Half\hL \dx^K = \hq \e^{2y(u)} \sums a_s Z_s{}^K Z_{sL}\dx^L
		         + F(u)  Z_{\o}{}^K Z_{\o\, L} \dx^L,
     \cm
	 F(u) = \frac{1}{Y_\o^4}
	 	\biggl[\frac{(y_u+k)^2}{d_0\beta_u}\biggr]_u,
\eeq
     where  $\hq = \hq_\o = \sums Q_s^2$ and
     $y(u)$ is given by (\ref{2-3}). As before, the
     perturbations $\vZ_{\bot}\delta\ox$ (where $\vZ_\bot$ belong to the
     plane $\oV_\bot$ orthogonal to all $\vZ_s$, whose dimension is
     $\dim\oV-m \geq 0$) are decoupled and obey the equation (\ref{W3}),
     giving no unstable modes.

     Multiplying (\ref{Wm}) by $\oZ_{\o}$, one obtains a decoupled equation
     for $\xi_{\o} = \oZ_{\o} \delta\ox$:
\beq                                                             \label{Wo}
     \hL \xi_{\o} = 2 U_{\o}(u) \xi_{\o}, \cm
        	      U_{\o}(u) = (\hq\e^{2y} + F) Z_{\o}^2
\eeq
     where $Z_{\o}^2 = \oZ_{\o}^2$.
     Since, as is directly verified, $U_{\o}(u) \geq 0$, this mode is also
     stable.

     The remaining $(m-1)$ degrees of freedom may be described in terms of
     the vectors $\oW_s = \oZ_s - \oZ_{\o}$ and the functions
     $\xi_s = \oW_s \delta\ox$, such that
\beq                                                              \label{m3}
	\oW_s \oZ_{\o} =0; \cm \sums a_s \oW_s =0;
						      \cm \sums a_s\xi_s=0.
\eeq
     Using (\ref{m2}) and (\ref{m3}), one obtains the following $m$
     equations, coupled due to (\ref{m3}), for $(m-1)$ independent variables:
\beq
     \hL \xi_s = 2\hq \e^{2y} \sum_{s'=1}^{m} K_{ss'}\xi_{s'},
     \inch
     		K_{ss'} = a_{s'}\oW_s \oW_{s'}.		        \label{Ws}
\eeq
     Excluding one of the unknowns, say, $\xi_m$, by
     virtue of (\ref{m3}), we arrive at a determined set of wave
     equations for $\eta_s = \xi_s - \xi_m$, $s = 1, \ldots, m-1$:
\beq
     \hL \eta_s = 2\hq \e^{2y} \sum_{s'=1}^{m-1}F_{ss'}\eta_s',\cm
     							       \label{Weta}
     F_{ss'} = a_{s'} \oW_{s'} \biggl(\oW_s
	    	+\frac{1}{a_m}\sum_{s''=1}^{m-1}a_{s''}\oW_{s''}\biggr).
\eeq
     This is a good way of studying specific models. In the general case,
     however, the situation looks more transparent if we
     consider, instead, an auxiliary system with $m$ independent unknowns,
     described by \eqs (\ref{Ws}) where $\oW_m$ is slightly shifted from
     its true value by some $\Delta \oW$, so that all $\oW_s$ become
     linearly independent; the relation among $\xi_s$ in (\ref{m3}) is then
     cancelled as well. Our system is restored when $\Delta\oW \to 0$.

     For the auxiliary system the matrix $(\oW_s \oW_{s'})$ is symmetric
     and positive-definite; if all $a_s$ are equal, the same is
     true for the matrix of coefficients in (\ref{Ws}), $(K_{ss'})$,
     hence there is a similarity transformation
     bringing it to a diagonal form with its positive eigenvalues along the
     diagonal.  Such a transformation applied to \eqs (\ref{Ws})
     decouples them into $m$ separate wave equations like (\ref{Wo}),
     with some positive function replacing $U(u)$. In the limit
     $\Delta\oW \to 0$, the worst thing that can happen is that some of the
     eigenvalues tend to zero, giving for some combinations of $\xi_s$ the
     equation $\hL\xi=0$ which, as we know, does not lead to an instability.
     One can assert ``by continuity'' that this picture is generic, at least
     for $a_s$ close enough to being equal, and stability is again
     concluded according to Proposition 2.

     On the contrary, when the numbers $a_m > 0$ are different, one cannot
     guarantee that the non-symmetric matrix $K_{ss'}$ is similar to a
     diagonal one \cite{gant}. A failure in its diagonalization can be
     connected with the occurrence of a pair (or pairs) of complex roots
     $\lambda_s$ of the characteristic equation $\det
     |K_{ss'}-\lambda\delta_{ss'}|=0$.  In this case there is at least one
     pair of coupled perturbations for which a special investigation is
     necessary.  An inspection of the characteristic equation shows that
     the matrix $K_{ss'}$ cannot have negative eigenvalues, therefore a
     separate unstable mode cannot occur and the only possible instability
     can be connected with coupling between modes.

     In particular, in an arbitrary OS BH solution there is a
     subfamily where all $y_s(u)$ coincide (i.e., the constants $u_s$ are
     the same for all $s$), so that the branes form a BOS block, and
     it turns out that all $a_s$ are also equal, as well as the squared
     charges $Q_s^2$. The above reasoning shows that such solutions are
     stable.

     If $\rank (K_{ss'}) < m-1$, that is, there are additional linear
     dependences among $\oZ_s$, then some combinations of $\xi_s$ decouple
     leading to equations of the form $\hL\xi=0$, and for the remaining
     modes the above discussion can be repeated with slight modifications.

     This is what can be said about the general case of single-block BOS BH
     solutions. If there is a block of only two branes ($m=2$), one can
     make a common stability conclusion generalizing the one made in
     \sect 6.1. Indeed, for $\xi_{\o} = a_1\xi_1 + a_2\xi_2$ there is
     \eq (\ref{Wo}), whereas for $\xi_- = \xi_1 - \xi_2$ one obtains
\beq
     \hL \xi_- = 2\hq \e^{2y} a_1 (1-a_1) (\oZ_1-\oZ_2)^2 \xi_-. \label{W-}
\eeq
     In the special case $Z_1^2=Z_2^2=Z^2$ one recovers (\ref{2-4}),
     (\ref{2-5}).

     For $m \geq 3$ one has to study specific models individually.

\section {Concluding remarks}

     We have shown that all static single-brane BH solutions with the metric
     (\ref{2.11}) are stable under linear \sph perturbations, whereas non-BH
     solutions possessing naked singularities of different types are
     unstable. Very probably other singular solutions, for which
     perturbation equations do not decouple, are unstable as well, since, as
     known from vibration theory, coupling between modes can hardly stabilize
     them. On the contrary, coupled modes can be unstable even when single
     ones are stable. It is therefore of interest to study the stability
     properties of more complex BH solutions; this work is in progress.

     We have also shown that the BH stability conclusion can be extended to
     some BHs with multiple intersecting \branes, namely, for the BOS case,
     characterized by a single function $y(u)$ (i.e., all $y_s$
     coincide). It turns out that for such backgrounds the wave equations
     for perturbations also generically decouple and the absence of unstable
     modes can be proved. Though, such a general proof is available only in
     two cases: (i) two-brane BH solutions ($m=2$) and (ii) equal-charge
     subfamilies of arbitrary OS solutions. Nontrivial brane systems with
     $m>2$ should be studied individually to see whether or not the
     corresponding matrix $(K_{ss'})$ in (\ref{Ws}) can be diagonalized over
     the field of real numbers.  If yes, the solution is stable, otherwise a
     further study of coupled modes is necessary.

     It should be stressed that a BOS-block solution exists for an {\it
     arbitrary\/} set of linearly independent vectors $\vY_s$. In
     particular, if any multi-brane static BH solution for a certain set
     of input parameters with independent vectors $\vY_s$ is known,
     e.g., any OS or BOS solution (see \sect 3), then the additional
     requirement that all the functions $y_s$ coincide selects from it a
     special BOS-block solution, for which a stability study can be
     performed as described above. The only restriction is the requirement
     $a_\mu(\o)>0$ for the charge factors obeying \eqs (\ref{*3}).

     Some technical points are worth mentioning.
     First, in gravitational stability studies it is sometimes rather hard
     to separate real physical perturbations from purely gauge degrees of
     freedom. We avoid this problem by obtaining the set of wave equations
     (\ref{W}) where the number of equations is precisely the number of
     dynamical degrees of freedom, represented by scalars in the physical
     space-time $\Mph$.  Due to the latter circumstance, one more
     complication is avoided: when dealing with vector and tensor
     perturbations of BHs, one has to take into account the apparent
     singularity of the metric on the horizon; to properly formulate the
     boundary conditions, it is then necessary to pass to Kruskal-like
     coordinates; to be admissible, and the perturbations are required to be
     finite on the future horizon \cite{vish, glaf1, glaf2}. In our case
     the perturbations are scalars, so the finiteness requirement can be
     imposed in any coordinates. The choice of gauge only remains important
     for making the treatment more transparent.

     To conclude, we would like to emphasize that our consideration did
     not depend on the number and dimensions of the factor spaces in the
     original space-time $\M$, on the number of scalar fields $\varphi^a$
     and on the particular values of their coupling constants $\lambda_{sa}$.

\section*{Appendix}
\renewcommand{\theequation}{A.\arabic{equation}}
\sequ 0

     Consider, for illustration, some solutions of $D=11$ supergravity,
     representing the low-energy limit of M-theory, as examples of systems
     to which our stability results apply.

     The action (\ref{2.1}) for this theory does not contain scalar fields
     ($\varphi^a = \lambda_{sa} =0$) and the only $F$-form is of rank 4,
     whose various nontrivial components $F_s$ (elementary $F$-forms
     according to \sect 2, to be called simply $F$-forms) are associated
     with electric 2-branes [for which $d(I_s)=3$] and magnetic 5-branes
     [such that $d(I_s)=6$] (see \cite{peter} and references therein). The
     orthogonality conditions (\ref{3.21}) are satisfied if the following
     intersection rules hold:
\beq
     3\cap 3 =1, \cm 3\cap 6=2, \cm 6\cap 6 = 4.            \label{A1}
\eeq
     (the notations are evident); for all $F$-forms $Y_s^2 =2$.

     We will designate the branes by figures labelling their world volume
     coordinates (covered by $I_s$), beginning with ``1'' which
     corresponds to the time axis. Thus, e.g., (123) is an electic 2-brane
     whose world volume includes the time axis $\M_1 = \R_t$ and two extra
     dimensions. The number of dimensions where branes can be located is
     $D-1-d_0 = 10 - d_0$.

\medskip\noi
     {\bf 1.}
     Single-brane BH solutions are described by (\ref{5.7}), (\ref{5.8})
     where all sums and products in $s$ consist of a single term. These
     solutions are well known; the metric (\ref{2.11}) can be presented as
\beq
     ds_{11}^2 = H^{d(I)/9}                                \label{ds11}
      \biggl[ -\frac{1-2k/(\od r^{\od})}{H}\,dt^2
              + \biggl(\frac{dr^2}{1-2k/(\od r^\od)} + r^2 d\Omega^2\biggr)
	      + H^{-1} ds_{\rm on}^2 + ds_{\rm off}^2
	      		\biggr]
\eeq
     where $H = H(r) = 1 + p/(\od r^\od)$, $p = \sqrt{k^2 + 2Q^2} - k$,
     $\od = d_0-1$; $ds_{\rm on}^2$ and $ds_{\rm off}^2$ are the
     ``on-brane'' and ``off-brane'' extra-dimension line elements,
     respectively; the dimension $d_0$ of the sphere $\M_0$ varies from 2 to
     7 for $d(I)=3$ (an electric brane) and from 2 to 4 for $d(I)=6$ (a
     magnetic brane). In particular, the cases of maximum $d_0$, when
     off-brane extra dimensions are absent, correspond in the extremal
     near-horizon limits to the famous structures $AdS_4 \times S^7$
     (electric) and $AdS_7 \times S^4$ (magnetic). All these solutions are
     stable under linear \sph perturbations.

\medskip\noi
     {\bf 2.}
     Some examples of orthogonal systems (OS), whose stability in the
     general case is yet to be studied, are:
\bear
     {\bf (i)} &&\quad \mbox{(123), (145), (167)
     			--- 3 electric branes; $d_0=2$ or 3.}
 \nn
     {\bf (ii)}&&\quad \mbox{(123), (124567) ---
     		1 electric and 1 magnetic branes; $d_0=2$ or 3.}
 \nn
     {\bf (iii)}&&\quad (123),\ (145),\ (124678),\ (135678);\ \ d_0=2.
\earn
     The metrics are easily found from (\ref{5.7}) with $D-2=9$, $Y_s^2=2$
     and the equalities marked $\eqos$.

     The systems (i) with $d_0=3$ and (iii) are remarkable in that their
     extremal limits have regular horizons and the near-horizon geometries
     are, respectively, $AdS_2 \times  S^3 \times T^6$ and
     $AdS_2 \times S^2 \times T^7$ if the remaining extra dimensions are
     compactified on tori.

     When the orthogonal systems form BOS-blocks (i.e., in the special case
     of equal charges and a unique function $y(u)$), the solutions are
     stable according to \sect 6.2.

\medskip\noi
     {\bf 3.}
     All two-brane BOS-block BHs are stable according to \sect 6.1,
     for instance,
\[
     {\bf (i)}\ \ (123),\ (123456); \ \ \vY_1 \vY_2 = 1. \cm\cm
     {\bf (ii)}\ \ (123),\ (145678); \ \ \vY_1 \vY_2 = -1.
\]
     The norms are equal ($Y_s^2=2$), and the angle $\theta$ is 60\deg in
     case (i) and 120\deg in case (ii).

\medskip\noi
     {\bf 4.}
     Many seemingly possible three-brane blocks turn out to be forbidden
     due to a zero value of a certain $a_s$ (see \sect 3.3). Consider,
     e.g.,
\beq
     (123),\ (145),\ (123678);\cm
     (\vY_s \vY_{s'}) = \pmatrix { 2 &  0 &  1 \cr
                                   0 &  2 & -1 \cr
				   1 & -1 &  2 \cr }.
\eeq
     One easily finds from (\ref{*3}) that
     $(a_1,\ a_2,\ a_3)= (0, 1/2, 1/2)$, so one of the charges should be
     zero, which means that such a system cannot exist.

\medskip\noi
     {\bf 5.}
     The following is an example of a single-block BH whose stability can be
     established by an individual study as described in \sect 6.2:
\beq
     (123),\ (145),\ (123467); \ \ d_0 =2\ {\rm or}\ 3;\cm  \label{A.4}
     (\vY_s \vY_{s'}) = \pmatrix { 2 &  0 &  1 \cr
                                   0 &  2 &  0 \cr
				   1 &  0 &  2 \cr }.
\eeq
     From (\ref{*3}) it follows
\beq
     (a_1,\ a_2,\ a_3)= (2/7,\ 3/7,\ 2/7); \cm
     \vY_\o = \sum_{s=1}^{3}a_s\vY_s;      \cm   \vY_\o^2 = 6/7.
\eeq
     Suppose for certainty $d_0=3$. Then, in agreement with (\ref{st9}),
\beq
    (\oZ_s \oZ_{s'}) =  \pmatrix{ 4/3  & -2/3 &  1/3 \cr
        			 -2/3  &  4/3 & -2/3 \cr
				  1/3  & -2/3 &  4/3 \cr},
     \cm Z_\o^2 = \frac{4}{21}.
\eeq
     The next stage is to separate the perturbation $\oZ_\o \delta\ox$,
     after which the remaining two degrees of freedom obey \eqs (\ref{Weta})
     of the form
\beq                                                          \label{A.7}
     \hL \eta_s = 2\hq \e^{2y} \sum_{s'=1}^{2}F_{ss'}\eta_{s'},
		  \cm F_{ss'} =
			       \pmatrix {\ 2/7  &  0    \cr
                                          -2/7  &  6/7  \cr}.
\eeq
     The characteristic equation $\det |F_{ss'}-\lambda\delta_{ss'}|=0$
     has the form $(2/7-\lambda)(6/7-\lambda)=0$, and, according to
     Proposition 2, the positivity of its roots proves the stability of the
     background configuration.

\Acknow
{V.M. is grateful to Dept. Maths., University of the Aegean, Greece, and to
Dept.  Phys., Nara University, Japan, for their hospitality during his stay
there in October and November-December 1999, respectively. K.B. acknowledges
the hospitality of the colleagues of DFis-UFES, Vit\'oria, ES, Brazil during
his stay there in November-December 1999.  The work was supported in part by
the Russian Basic Research Foundation and by the Russian Ministry of Science
and Technologies.}

\end{document}